\documentclass[manuscript]{acmart}

\usepackage{arydshln}
\usepackage{subcaption}
\usepackage{amsmath}
\usepackage{booktabs} 
\usepackage{xspace}

\AtBeginDocument{%
  \providecommand\BibTeX{{%
    \normalfont B\kern-0.5em{\scshape i\kern-0.25em b}\kern-0.8em\TeX}}}

\setcopyright{rightsretained}

\newcommand{\modelvae}{\texttt{VAE}\xspace}
\newcommand{\modelrand}{\texttt{RAND}\xspace}
\newcommand{\modelslim}{\texttt{SLIM}\xspace}
\newcommand{\modelals}{\texttt{ALS}\xspace}
\newcommand{\modelbpr}{\texttt{BPR}\xspace}
\newcommand{\modelpop}{\texttt{POP}\xspace}
\newcommand{\modelitemknn}{\texttt{ItemKNN}\xspace}

\begin{document}

\copyrightyear{2021} 
\acmYear{2021} 
\acmConference[RecSys '21]{Fifteenth ACM Conference on Recommender Systems}{September 27-October 1, 2021}{Amsterdam, Netherlands}
\acmBooktitle{Fifteenth ACM Conference on Recommender Systems (RecSys '21), September 27-October 1, 2021, Amsterdam, Netherlands}
\acmDOI{10.1145/3460231.3478843}
\acmISBN{978-1-4503-8458-2/21/09}
\title[Gender-specific Popularity Bias of Music Recommendation Algorithms]{Analyzing Item Popularity Bias of Music Recommender Systems: Are Different Genders Equally Affected?}




 \author{Oleg Lesota}
 \email{oleg.lesota@jku.at}
 \affiliation{
 \institution{Johannes Kepler University Linz 
 and Linz Institute of Technology
 }
 \country{Austria}
 }

 \author{Alessandro B. Melchiorre}
 \email{alessandro.melchiorre@jku.at}
 \affiliation{
 \institution{Linz Institute of Technology} 
 \country{Austria}
 }

 \author{Navid Rekabsaz}
 \email{navid.rekabsaz@jku.at}
 \affiliation{
 \institution{Johannes Kepler University Linz 
 and Linz Institute of Technology} 
 \country{Austria}
 }

 \author{Stefan Brandl}
 \email{stefan.brandl@jku.at}
 \affiliation{
 \institution{Johannes Kepler University Linz}
 \country{Austria}
 }
 
 \author{Dominik Kowald}
 \email{dkowald@know-center.at}
 \affiliation{
 \institution{Know-Center GmbH}
 \country{Austria}
 }

 \author{Elisabeth Lex}
 \email{elisabeth.lex@tugraz.at}
 \affiliation{
 \institution{Graz University of Technology}
 \country{Austria}
 }
 
 \author{Markus Schedl}
 \authornote{This is the corresponding author.}
 \email{markus.schedl@jku.at}
 \affiliation{
\institution{Johannes Kepler University Linz 
and Linz Institute of Technology} 
 \country{Austria}
 }


\renewcommand{\shortauthors}{Lesota et al.}

\begin{abstract}
Several studies have identified discrepancies between the popularity of items in user profiles and the corresponding recommendation lists. Such behavior, which concerns a variety of recommendation algorithms, is referred to as popularity bias. Existing work predominantly adopts simple statistical measures, such as the difference of mean or median popularity, to quantify popularity bias. Moreover, it does so irrespective of user characteristics other than the inclination to popular content. 
In this work, in contrast, we propose to investigate popularity differences (between the user profile and recommendation list) in terms of median, a variety of statistical moments, as well as similarity measures that consider the entire popularity distributions (Kullback-Leibler divergence and Kendall's $\tau$ rank-order correlation). This results in a more detailed picture of the characteristics of popularity bias.
Furthermore, we investigate whether such algorithmic popularity bias affects users of different genders in the same way. 
We focus on music recommendation and conduct experiments on the recently released standardized LFM-2b dataset, containing listening profiles of Last.fm users. We investigate the algorithmic popularity bias of seven common recommendation algorithms (five collaborative filtering and two baselines).
Our experiments show that (1) the studied metrics provide novel insights into popularity bias in comparison with only using average differences, (2) algorithms less inclined towards popularity bias amplification do not necessarily perform worse in terms of utility (NDCG), (3) the majority of the investigated recommenders intensify the popularity bias of the female users.

\end{abstract}




\maketitle

\section{Introduction}\label{sec:intro}
Popularity bias in recommender systems refers to a disparity of item popularities in the recommendation lists. Most commonly, this means that a disproportionally higher number of popular items than less popular ones are recommended~\cite{ekstrand2018all}.
The existence of such a popularity bias 
has been evidenced in different domains already, e.g., movies~\cite{DBLP:conf/recsys/AbdollahpouriMB19}, music~\cite{DBLP:conf/ecir/KowaldSL20}, or product reviews~\cite{abdollahpouri2017controlling}.
Collaborative filtering recommenders are particularly prone to popularity biases because the data they are trained on 
already exhibit an imbalance towards popular items, i.e., more user--item interactions are available for popular items than less popular ones~\cite{abdollahpouri2019managing}.

The distribution of item popularities in most domains, in particular in the music domain, which we target in this work, shows a long-tail characteristic~\cite{DBLP:books/daglib/0025137}.
A recommendation algorithm introduces no further \textit{algorithmic bias} when the distribution of popularity values of recommended items (tracks) exactly matches the distribution of popularity values of already consumed items (listening history) for each user.

We identify two shortcomings of existing studies of popularity bias:
First, popularity bias is commonly quantified using simple statistical aggregation metrics, predominantly comparing arithmetic means computed on some count of the user--item interactions~\cite{DBLP:conf/recsys/AbdollahpouriMB19,DBLP:conf/ecir/KowaldSL20}. These are not robust against outliers often present in music listening data.
Second, popularity bias is typically studied irrespective of user characteristics. Therefore, the extent to which users of different groups (e.g., age, gender, or cultural background) are affected remains unclear. 
We set out to approach these shortcomings in the music domain by posing the following research questions:


\begin{itemize}
\item \textit{RQ1: Which novel insights into popularity bias can be obtained by quantifying algorithmic popularity bias based on the median, a variety of statistical moments, and similarity measures between popularity distributions?} 
\item \textit{RQ2: Do algorithmic popularity biases affect users of different genders in the same way?} 
\end{itemize}

We find that users of different genders are affected by algorithm-inflected bias differently, such that the majority of the models expose female users to more biased results. Also, algorithms less inclined towards popularity bias amplification do not necessarily perform worse in terms of utility (NDCG). Finally, the studied metrics provide novel insights into popularity bias in comparison with only using average differences.


\section{Related Work}\label{sec:related}


We focuse on popularity bias, a well-studied form of bias in recommender systems research. This form of bias refers to the underrepresentation of less popular items in the produced recommendations and can lead to a significantly worse recommendation quality for consumers of long tail or niche items~\citep{DBLP:conf/ecir/KowaldSL20,lex2020modeling,DBLP:conf/recsys/AbdollahpouriMB19,Jannach:2015:BHH:2792838.2800182}.
\citet{DBLP:conf/recsys/AbdollahpouriMB19} show that state-of-the-art movie recommendation algorithms suffer from popularity bias, and introduce the delta-GAP metric to quantify the level of underrepresentation. As shown in \citet{DBLP:conf/ecir/KowaldSL20}, in particular users interested in niche, unpopular items suffer from a worse recommendation quality. 
The authors use the delta-GAP metric in the domain of music recommendations, and find that the delta-GAP metric does not show a difference between ``niche'' and ``mainstream'' users. The reason for this could be that a group-based metric is not suitable for the complexity of music styles, as user groups can be quite diverse within themselves~\citep{kowald2021support}.
\citet{Zhu} address a related problem of item under-recommendation bias, expressing it with ranking-based statistical parity and ranking-based equal opportunity metrics. \citet{Boratto_2021} propose metrics quantifying the degree to which a recommender equally treats items along the popularity tail.

In contrast to these works, 
we study differences between popularity distributions of consumed and recommended items for each user. We express them in terms of the median as well as several statistical moments and similarity measures. In addition, we combine research strands on popularity bias and gender bias by analyzing how female and male listeners are affected by popularity bias.


\section{Measuring Popularity Bias}\label{sec:methods}
 We introduce ways to express popularity bias as quantified dissimilarity between popularity distributions of recommended and consumed items for each user.

\begin{figure*}[t]
\centering
\subcaptionbox{\label{fig:distr_examples}}{\includegraphics[width=0.49\textwidth]{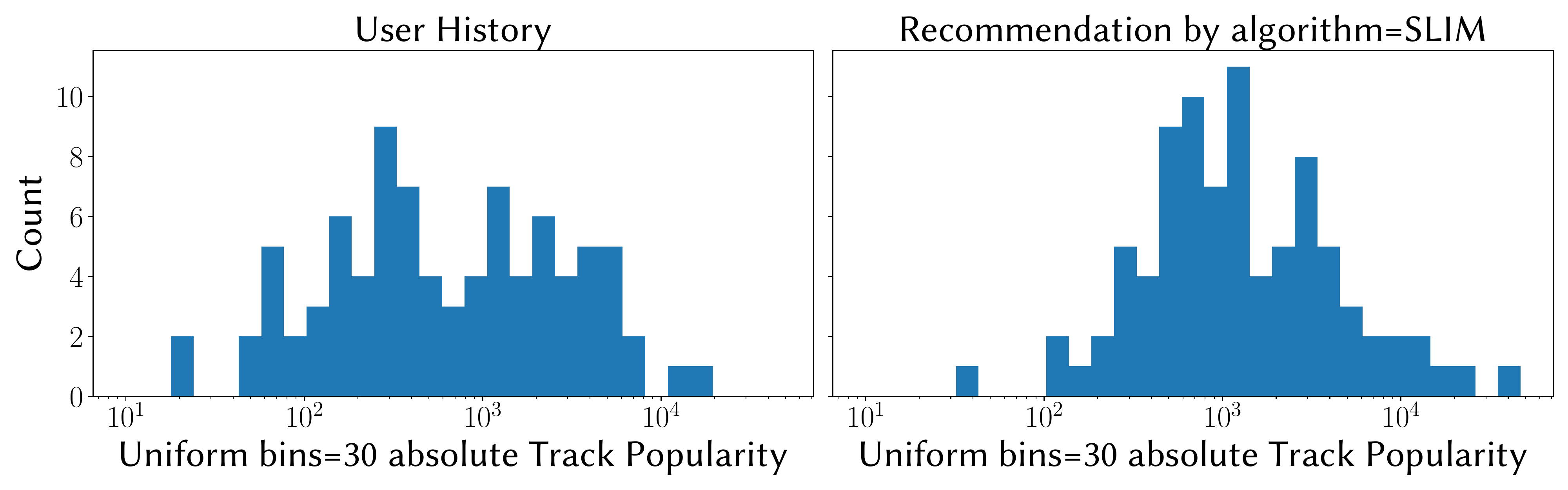}}
\subcaptionbox{\label{fig:distr_examples_decBinned}}{\includegraphics[width=0.49\textwidth]{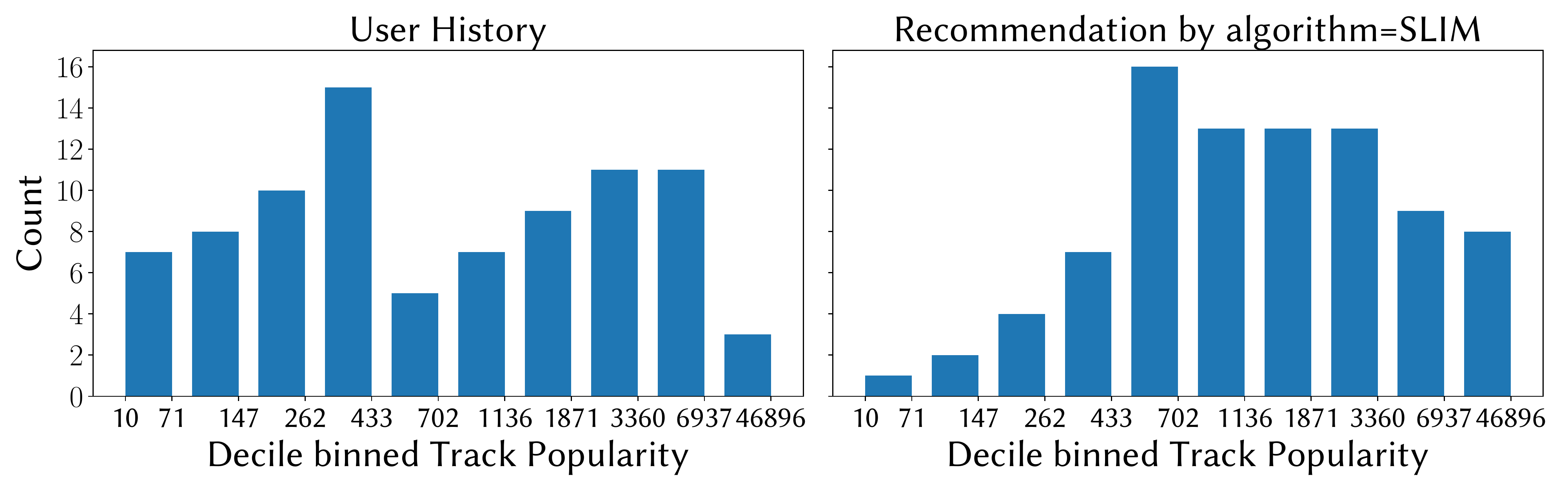}}

\caption{(a) shows equally binned (for visualization purposes only) distributions of popularity over the listening history (left) and the recommendation list (right) for the same user. On $x$-axis evenly binned popularity, on $y$-axis number of tracks in the distribution, falling into each bin. (b) demonstrates the same distributions binned with respect to the popularity distribution in the whole collection. This binning is employed for $KL$ and Kendall's $\tau$ calculations.}
\vspace{-0.3cm}
\end{figure*}

\subsection{Track Popularity Distributions}
We define $P(t)$ popularity of a track $t$ as the sum of its play counts over all users $u_i \in U$ in the dataset, namely $P(t) = \sum_{u_i \in U}{PC(t, u_i)}$. We then use these popularity estimates to derive the popularity distribution over each user's listening history and recommendation list. In order to make the popularity distribution $H_{u_i}(t)$ over a user's listening history $T_{hist}(u_i)$ comparable to the respective distribution $R_{u_i}(t)$ over the recommendation lists, we consider only the top of the recommendation list $T_{top\_rec}(u_i)$ so that its length (number of tracks) matches the length of the user's listening history $|T_{top\_rec}(u_i)| = |T_{hist}(u_i)|$. Therefore, we define the popularity distribution over the listening history and the recommendation list of user~$u_i$ as follows:
\vspace{-0.2cm}
\begin{align}
H_{u_i}(t) = 
\begin{cases}
            P(t)  |  t \in T_{hist}(u_i) \\
            0  |  t \notin T_{hist}(u_i) \\
\end{cases}
&
R_{u_i}(t) = 
\begin{cases}
            P(t)  |  t \in T_{top\_rec}(u_i) \\
            0  |  t \notin T_{top\_rec}(u_i) \\
\end{cases}
\end{align}

To gain a better understanding of these distributions, Figure \ref{fig:distr_examples} shows an example of popularity distributions over a user's listening history $T_{hist}(u_i)$ and the corresponding recommendation list $T_{top\_rec}(u_i)$ produced by the SLIM recommender algorithm. 

\subsection{Metrics}
\subsubsection{Delta Metrics of Popularity Bias}
In order to measure the differences between these distributions, we first introduce a series of delta metrics to calculate the discrepancies between the listening history and recommendation list popularity distributions of each user, and then aggregate them to achieve per-system results. We study five $\% \Delta \mathcal{M}$ (percent delta) metrics where the metric $\mathcal{M}$ is one of the following: $Mean$, $Median$, $Variance$, $Skew$, $Kurtosis$. If $\mathcal{M}(H_{u_i}(t))$ and $\mathcal{M}(R_{u_i}(t))$ are the results of application of the same metric $\mathcal{M}$ to the two respective distributions, the respective $\% \Delta\mathcal{M}$ for the user $u_i$ is calculated as:
$\%\Delta\mathcal{M}_{u_i} = \frac{\mathcal{M}(R_{u_i}(t)) - \mathcal{M}(H_{u_i}(t))}{\mathcal{M}(H_{u_i}(t))} \cdot 100
\label{eq:delta_m}$

Positive $\%\Delta Mean$ and $\%\Delta Median$ indicate that overall more popular tracks are recommended to the user. Since $Mean$ is sensitive to outliers, the interplay between these metrics provides additional information about the changes in popularity. Positive $\%\Delta Variance$ means that the list of recommended items is more diverse in terms of different popularity values than the user's history. This can also mean an increase in bias towards more popular items, as the most popular items are sparsely distributed across the popularity range.
Positive $\%\Delta Skew $ denotes that the right tail of the recommendation list distribution is heavier (with respect to the left tails) than the one belonging to the user-history distribution. A positive value therefore means that more items tend to have lower popularity from the range of the distribution.
Finally, positive $\%\Delta Kurtosis $ shows that the tails of the recommended tracks' popularity distribution are heavier than of its counterpart, and the distribution itself is in a way closer to uniform distribution. 

Finally, the discussed metrics describe the difference between the distributions for a particular user. In order to represent the change across all users, we take the median of the per-user values. 

\subsubsection{Kullback–Leibler Divergence and Kendall's $\tau$ as Measures for Popularity Bias}
In order to compare the entire popularity distributions, we utilize Kullback–Leibler Divergence ($KL$) and Kendall's $\tau$ ($KT$). For each user, we apply these metrics to the corresponding $H_{u_i}(t)$ and $R_{u_i}(t)$ decile-binned with respect to the popularity distribution over the whole collection ($P(t)$). The bins are chosen in such a way that the cumulative popularity of all tracks of the collection belonging into one bin constitutes approximately $10\%$ of the total popularity of all tracks of the whole collection. Figure~\ref{fig:distr_examples_decBinned} shows the distributions from Figure~\ref{fig:distr_examples} binned this way. In our dataset, the bin corresponding to the most popular tracks is constituted by only $161$ items whose popularity ranges from about 7k to 47k total play counts. Each bin covers items that are roughly half as popular as the next decile bin and two times as popular as the previous decile bin. Such binning allows the two metrics to be less sensitive to minor differences between the distributions and concentrate on the shifts between different popularity categories.

$KL$ estimates the dissimilarity of two 
distributions, in our case, between the user's listening history and recommendation list popularity distributions. It is defined as
\vspace{-0.2cm}
\begin{equation}
    KL_{u_i}(\hat{H}_{u_i}(b) | \hat{R}_{u_i}(b)) =  \sum_{b_j \in B}{\hat{H}_{u_i}(b_j)\log{\frac{\hat{H}_{u_i}(b_j)}{\hat{R}_{u_i}(b_j)}}}
    \vspace{-0.2cm}
\end{equation}

where $\hat{H}_{u_i}(b)$ and $\hat{R}_{u_i}(b)$ are decile-binned and normalized versions of the distributions and $b_j \in B$ represent the ten bins. $KL$ compares the two distributions and increases with every mismatch in the item counts. It is particularly sensitive to the case when for a bin the user gets recommended fewer tracks than they have in their listening history.



While $KL$ Divergence is sensitive to actual count changes, Kendall's $\tau$ metric reflects whether the order of bins is the same for the two distributions when ranked according to the respective counts. Kendall's $\tau$ is calculated as $KT_{u_i}(\hat{H}_{u_i}(b), \hat{R}_{u_i}(b)) =  \frac{C - D}{C + D}$, where $C$ represents the number of pairs of bins that have the same respective ranking in both distributions (concordant pairs) and $D$ the number of pairs of bins that have the different respective ranking in the two distribution (discordant pairs). For example, looking at  Figure~\ref{fig:distr_examples_decBinned}, the first two bins are concordant ($\in C$) as in both cases, more items fall into the second bin. While the first and the last bins are discordant ($\in D$) as in the listening history distribution, the first bin has more items. However, the recommended distribution shows the opposite. This way, $KT$ shows whether there are common patterns (correlations) in the two distributions, and it reaches its maximum value of $1$ when the two distributions are identical from the bin-ranking point of view. 
Similar to $\%\Delta\mathcal{M}$ metrics, we use the median of the per-user values to measure the differences across all users for $KL$ and $KT$. 

\begin{table}[t]
\centering
\caption{\label{tab:exp_data_stats} Statistics of the dataset. Number of Users, Tracks 
and listening events (LEs) are reported across F(emale) and M(ale) separately and also together (All). Mean and standard deviation (indicated after $\pm$) of the interactions of users with tracks 
and LEs are indicated in the last three columns, respectively.}
\scalebox{0.9}{
\begin{tabular}{l|rrrrrrrr}
\toprule
Gender & \multicolumn{1}{c}{Users} & \multicolumn{1}{c}{Tracks} 
& \multicolumn{1}{c}{LEs} & \multicolumn{1}{c}{Tracks/User} 
& \multicolumn{1}{c}{LEs/User} \\
\midrule
 All & $19,972$ & $99,831$ 
 & $19,906,272$ & $142 \pm 172$ 
 & $997 \pm 1,571$ \\
 
 F & $4,415$ & $70,980$ 
 & $3,397,310$ & $101 \pm 121$ 
 & $769 \pm 1,158$ \\
 
 M & $15,557$ & $99,810$ 
 & $16,508,962$ & $153 \pm 182$ 
 & $1,061 \pm 1,664$ \\
\bottomrule
\end{tabular}
}
\vspace{-0.2cm}
\end{table}

\section{Experiment Setup}\label{sec:setup}
\subsection{Recommendation Algorithms}
To study algorithmic popularity biases, we examine different commonly used collaborative filtering algorithms 
(i.e., heuristic, neighborhood based, matrix factorization, and autoencoders)~\cite{dacrema2019we,MELCHIORRE2021102666}:
\begin{itemize}
    \item Random Item (\modelrand): A baseline algorithm that recommends for each user random items. It avoids recommending already consumed items. 
    \item Most Popular Items (\modelpop): A baseline that implements a heuristic-based algorithm that recommends the same set of overall most popular items to each user.
    \item Item k-Nearest Neighbors (\modelitemknn) \cite{deshpande2004item}: A neighborhood-based algorithm that recommends items based on item-to-item similarity. Specifically, an item is recommended to a user if the item is similar to the items previously selected by the user. \modelitemknn uses     statistical measures to compute the item-to-item similarities.
    \item Sparse Linear Method (\modelslim) \cite{ning2011slim}: Also a neighborhood-based algorithm, but instead of using predefined similarity metrics, the item-to-item similarity is learned directly from the data with a regression model.
    \item Alternating Least Squares (\modelals) \cite{hu2008collaborative}: A matrix factorization approach that learns user and item embeddings such that the dot product of these two approximates the original user-item interaction matrix.
    \item Matrix factorization with Bayesian Personalized Ranking (\modelbpr) \cite{rendle2012bpr}: Learns user and item embeddings, however, with an 
    optimization function that aims to 
    rank the items consumed by the users according to their preferences (hence, personalized ranking) instead of predicting the rating for a specific pair of user and item.
    \item Variational Autoencoder (\modelvae) \cite{liang2018variational}: An autoencoder-based algorithm that, given the user's interaction vector, estimates a probability distribution over all the items using a variational autoencoder architecture.
\end{itemize}

For training the models, we use the same hyperparameter settings as provided by \citet{MELCHIORRE2021102666}.


\begin{table*}[t]
\begin{center}
\caption{\label{tbl:results_all_fm} Results of algorithm-inflected popularity bias evaluation 
in terms of the seven introduced metrics and NDCG@10. Each model is represented by three rows. The row $All$ gives the results on the whole dataset. The rows $\Delta Female$ and $\Delta Male$ describe the difference in the result between the user group and the whole population 
in the dataset. For example, the $\%\Delta Variance$ for algorithm \modelslim for $All$ 
of $56.0$ 
denotes 
a median increase in popularity variance (between listening history and recommended list) 
of $56\%$ over all users. The corresponding $\Delta Female$ value of $ -17.4$ means that the variance increase for this group 
is $56.0 - 17.4 = 38.6\%$.}
\scalebox{0.95}{
\begin{tabular}{ c c| r r r r r|r r|r } 
\toprule
Alg. & Users &  $\%\Delta Mean$  &  $\%\Delta Median$ & $\%\Delta Var.$ & $\%\Delta Skew$    & $\%\Delta Kurtosis$  & $KL$   & Kendall's $\tau$ & NDCG@10\\ \midrule

 & $All$ & $-91.8$ & $-87.2$ & $-99.5$ & $11.5$ & $15.3$ & $3.904$ & $0.165$ & $0.000$ \\ \cdashline{2-10}
RAND & $\Delta Female$ & $-1.8$ & $-3.5$ & $-0.2$ & $+0.0$ & $-3.5$ & $+0.976$ & $-0.189$ & $-0.000$ \\
 &  $\Delta Male$ & $+0.5$ & $+1.1$ & $+0.1$ & $-0.0$ & $+1.3$ & $-0.281$ & $+0.053$ & $+0.000$ \\ \hline 

 & $All$ & $432.5$ & $975.2$ & $487.0$ & $-58.0$ & $-87.0$ & $6.023$ & $0.057$ & $0.045$ \\ \cdashline{2-10}
POP & $\Delta Female$ & $+11.0$ & $+282.1$ & $-172.2$ & $-2.1$ & $-1.9$ & $+1.626$ & $-0.033$ & $+0.003$ \\
 & $\Delta Male$ & $-2.8$ & $-115.8$ & $+55.9$ & $+0.5$ & $+0.5$ & $-0.380$ & $+0.016$ & $-0.001$ \\ \hline 

 & $All$ & $121.8$ & $316.6$ & $72.6$ & $-25.2$ & $-43.9$ & $4.368$ & $0.046$ & $0.184$ \\ \cdashline{2-10}
ALS & $\Delta Female$ & $+9.9$ & $+27.4$ & $-7.1$ & $-3.2$ & $-5.4$ & $+0.467$ & $+0.110$ & $-0.017$ \\
 & $\Delta Male$ & $-2.7$ & $-6.6$ & $+1.6$ & $+0.8$ & $+1.5$ & $-0.121$ & $-0.023$ & $+0.005$ \\ \hline 

 & $All$ & $-49.0$ & $-3.7$ & $-87.4$ & $-14.8$ & $-29.4$ & $1.202$ & $0.268$ & $0.129$ \\ \cdashline{2-10}
BPR & $\Delta Female$ & $+5.2$ & $+7.7$ & $+2.1$ & $-1.4$ & $-3.9$ & $+0.476$ & $-0.043$ & $-0.011$ \\
 & $\Delta Male$ & $-1.1$ & $-1.9$ & $-0.6$ & $+0.4$ & $+1.1$ & $-0.110$ & $+0.010$ & $+0.003$ \\ \hline 

 & $All$ & $9.6$ & $4.6$ & $5.7$ & $-14.3$ & $-29.0$ & $0.175$ & $0.423$ & $0.301$ \\ \cdashline{2-10}
ItemKNN & $\Delta Female$ & $+2.0$ & $+5.8$ & $-2.6$ & $-2.1$ & $-3.2$ & $+0.128$ & $-0.037$ & $-0.042$ \\
 & $\Delta Male$ & $-0.5$ & $-1.3$ & $+0.9$ & $+0.8$ & $+0.9$ & $-0.020$ & $+0.008$ & $+0.012$ \\ \hline 

 & $All$ & $49.8$ & $99.8$ & $56.0$ & $-12.5$ & $-26.0$ & $0.424$ & $0.189$ & $0.365$ \\ \cdashline{2-10}
SLIM & $\Delta Female$ & $-6.4$ & $-13.1$ & $-17.4$ & $-1.7$ & $-4.6$ & $+0.217$ & $+0.052$ & $-0.048$ \\
 & $\Delta Male$ & $+1.9$ & $+3.9$ & $+5.6$ & $+0.6$ & $+1.1$ & $-0.029$ & $-0.012$ & $+0.014$ \\ \hline 

 & $All$ & $303.9$ & $736.3$ & $351.0$ & $-45.2$ & $-70.1$ & $4.823$ & $-0.028$ & $0.191$ \\ \cdashline{2-10}
VAE & $\Delta Female$ & $+10.1$ & $+56.4$ & $-69.3$ & $-6.2$ & $-6.6$ & $+0.633$ & $+0.146$ & $-0.020$ \\
 & $\Delta Male$ & $-2.3$ & $-20.4$ & $+17.3$ & $+1.8$ & $+2.1$ & $-0.161$ & $-0.042$ & $+0.006$ \\ \hline

\end{tabular}
}
\end{center}
\vspace{-0.4cm}
\end{table*}
\vspace{-0.2cm}

\subsection{Dataset and Evaluation Protocol}
We perform experiments on \textit{LFM-2b-DemoBias}~\cite{MELCHIORRE2021102666}, a subset of the \textit{LFM-2b} dataset\footnote{\url{http://www.cp.jku.at/datasets/LFM-2b}}.
As in~\cite{MELCHIORRE2021102666}, we only consider user-track interactions with a playcount (PC) > 1, possibly avoiding using spurious interactions likely introduced by noise. Furthermore, we only consider tracks listened to by at least 5 different users and, likewise, only users who listened to at least 5 different tracks. Moreover, we only consider listening events within the last 5 years, letting us focus more on possible popularity biases in the recent years. Lastly, we consider binary user-track interactions, i.e., 1 if the user has listened to the track at least once, 0 otherwise. 

The procedure described above results in a subset of 23k users over 1.6 million items. We finalize data preparation by sampling 100k tracks uniformly-at-random, which ensures that tracks of different popularity levels are equally likely to be included in the final dataset.
The statistics of the final dataset are reported in Table~\ref{tab:exp_data_stats}. We 
find that males represent the majority group in the dataset and that they 
create $\sim80\%$ of all listening events. 


As evaluation protocol, we employ a user-based split strategy \cite{liang2018variational,marlin2004collaborative}, 
i.e., we split the 19,972 users in the dataset 
into train, validation, and test user groups via a 60-20-20 ratio split. We carry out 5-fold cross validation and change these user groups in a round-robin fashion. 
The users in the training set and all their interactions are used to train the recommendation algorithms.
For testing and validation, we follow standard setups \cite{liang2018variational,DBLP:conf/www/Steck19} 
and randomly sample 80\% of the users' items as input for the recommendation models and use the remaining 20\% 
to calculate the evaluation metric.


\section{Results and Discussion}\label{sec:results}
The results are shown in Table~\ref{tbl:results_all_fm}. Each value in the \emph{All} rows, regarding the popularity bias metrics, shows the median value of the distribution of a given metric over all users. For instance, $\%\Delta Var.$ of 72.6\% for ALS denotes that the median increase in popularity variance is 72.6 percent between user's listening history and items recommended to each user across all users. \modelslim $KL$ 1.66 expresses that the median difference between user history popularity distributions and the corresponding recommended tracks popularity distributions is 1.66 in terms of $KL$ Divergence. The reported results regarding the genders indicate the changes in values in respect to the \emph{All} values.


Both baseline algorithms (\modelrand and \modelpop) show poor results on accuracy metrics. Notably, on the $\%\Delta$ popularity bias metrics, they show divergent behavior. Decreasing of $\%\Delta$ metrics of $Mean$, $Median$, $Variance$ and increasing of $Skew$ and $Kurtosis$ indicate that \modelrand provides a list of tracks whose popularity distribution is closer to uniform than those from users' listening histories. \modelpop has 
an opposite trend, as the recommended tracks' popularity distribution has a more pronounced peak, is skewed, and shifted towards more popular items. 
It also shows a substantial median increase of variance in popularity, which 
can be explained by the fact that in our dataset, 
the most popular tracks are sparsely distributed across a wide 
range of popularity values ($161$ track in the popularity range between 7k and 47k of total play counts). 
Thus, recommending tracks from this category leads to a high variance. High values for $KL$ for both baselines also indicate that the overall popularity distributions of the recommended items are highly different from those of the users' listening histories. The random recommender demonstrates a higher median Kendall's $\tau$, which means that its output better correlates with users' histories in terms of popularity distribution. 
Both 
neighborhood-based models (i.e., \modelitemknn and \modelslim) show a high performance in terms of NDCG and a moderate popularity bias in their recommendations according to the $\%\Delta$ metrics, which is lower 
compared to \modelvae and \modelals. In particular, \modelslim shows higher value in $\%\Delta$ $Mean$ and $Median$ compared to \modelitemknn, suggesting that the item-to-item similarities learned by \modelslim favors more popular items in the recommendations. 
\modelitemknn displays lower $KL$ and higher Kendall's $\tau$ than \modelslim, which means that its results better approximate users' listening histories 
(we attribute this to \modelitemknn being less sensitive to bias in the data as it does not require trainable parameters). These observations regarding the performance of the models indicate that a decrease in popularity bias does not necessarily lead to a significant performance drop. 
Comparing \modelals with \modelbpr, we can observe an opposite behavior. While providing less biased results, \modelbpr shows the poorest performance among all non-baseline algorithms. 
While \modelvae is similarly biased in terms of all metrics as \modelpop, it achieves a higher performance according to NDCG.

Comparing metrics between the two gender groups, we note that $\%\Delta$ $Mean$ and $Median$ is higher for female users.
That means that their recommendations 
contain more popular items and/or items of higher popularity than the ones they usually 
listen to, and for this user group, that 
effect is more pronounced (hence larger values). 
Considering 
that $\%\Delta Variance$ is 
lower for the female users, we conclude that their recommendations 
are less diverse in terms of track popularity while consisting of more popular items. Judging by $\%\Delta$ $Skew$, $Kurtosis$ as well as Kendall's $\tau$, we can suggest that most recommender algorithms provide recommendations with comparable popularity distributions 
to both male and female users. At the same time, a slightly larger $KL$ may mean a larger 
shift towards popular items for female users.
\modelitemknn 
is the least biased algorithm in our study. It features low absolute values of $\%\Delta$ $Mean$, $Median$ and $Variance$, meaning that its 
recommendations 
consist of tracks comparable to the user's listening history in terms of average popularity and variety. High Kendall's $\tau$ means that the shape of the popularity distribution of the recommendations 
best matches the user's history 
among all tested algorithms. Still, it is slightly biased towards more popular items, as shown by negative $\%\Delta$ $Skew$ and $KL$ (which combined with high Kendall's $\tau$ signalizes about a shift of the distribution). 



\section{Conclusions and Future Direction}\label{sec:concusion}
In this paper, we examine to what extent various music recommender systems amplify item popularity bias. 
We study seven metrics of popularity bias deviation and analyze the results of seven recommender algorithms for users of different genders and for the overall population 
in the dataset. 
Addressing \textit{RQ1}, we observe that the studied metrics 
capture considerably different aspects of difference between popularity distributions of consumed and recommended items. While $\%\Delta Mean$ and $\%\Delta Median$ 
tell us about overall trends (are recommended tracks more or less popular than consumed ones), $\%\Delta Variance$ expresses the change in the diversity between listening histories and recommendation lists, and $\%\Delta Skew$ and $\%\Delta Kurtosis$ hint on the difference of shapes between the two distributions. Finally, $KL$ Divergence and Kendall's $\tau$ allow insight into how well the distributions match on a more granular level. With regard to \textit{RQ2}, we found that while the investigated algorithms display various levels of popularity bias, the majority of them (\modelvae, \modelitemknn, \modelbpr, \modelals) expose the female users to more popularity biased results. In the future, we will approach mitigating model-imposed popularity bias, e.g., through adversarial training or incorporating bias into the loss function of the recommenders, as well as finding more expressive metrics describing differences in the popularity distributions. Additionally, we plan to split our users into groups according to mainstreaminess as in~\cite{DBLP:conf/ecir/KowaldSL20} to  
compare our metrics with the group-based delta-GAP metric used in that work.

\begin{acks}
This work was funded by the H2020 project AI4EU (GA: 825619), the Austrian Science Fund (FWF): P33526, and the FFG COMET program.
\end{acks}


\bibliographystyle{ACM-Reference-Format}
\bibliography{recsys_lbr_2021}

\end{document}